\documentstyle{article}
\newcounter{bla} 
\newenvironment{refnummer}{%
\list{[\arabic{bla}]}%
{\usecounter{bla}%
 \setlength{\itemindent}{0pt}%
 \setlength{\topsep}{0pt}%
 \setlength{\itemsep}{0pt}%
 \setlength{\labelsep}{2pt}%
 \setlength{\listparindent}{0pt}%
 \settowidth{\labelwidth}{[9]}%
 \setlength{\leftmargin}{\labelwidth}%
 \addtolength{\leftmargin}{\labelsep}%
 \setlength{\rightmargin}{0pt}}}
 {\endlist}

\def\ketm#1{  \left\vert  #1   \right\rangle   }

\def\tmm#1{      \buildrel - \over #1  }  
  
\def\tpm#1{      \buildrel + \over #1  }

\makeatletter

\def\AmS{{\protect\the\textfont\tw@
  A\kern-.1667em\lower.5ex\hbox{M}\kern-.125emS}}
\makeatother

\begin{document}

\title{THE PROGRAM FOR THE TRANSFORMATION OF ATOMIC STATE FUNCTIONS FROM $LS$- TO $jj$-COUPLED BASIS\\[1.0cm]}

\author{ Tomas \v{Z}alandauskas and Gediminas Gaigalas\\[1.0cm]
        {\em Vilnius University Research Institute of Theoretical Physics and Astronomy,}\\
        {\em A. Go\v{s}tauto 12, Vilnius 2600, LITHUANIA}}

\maketitle

\begin{abstract}

Here we present the program for the transformation of the expansions of the atomic state functions from 
$LS$- to $jj$-coupled bases.
The program is a part of the coupling optimization package \textsc{Laris} and allows to transform the 
multiconfigurational expansions with the arbitrary number of open shells (including $f$-electrons) 
and to identify atomic levels and classify the atomic spectra by means of intermediate 
quantum numbers of $LS$- and $jj$-couplings.

\end{abstract}

\vspace{6cm} PACS Ref.: 31.15Ar, 31.15.-p, 02.70.-c.

\vspace{0.3cm}
{\em Keywords:} {\small atomic structure, $LS$-coupling, $jj$-coupling,
$LS-jj$ transformation, optimal coupling scheme, classification of atomic spectra.}

\newpage

{\bf PROGRAM SUMMARY} \\

\begin{small}


\noindent {\em Title of program:}  \textsc{Lsjj} \\[10pt]
%
%
{\em Program obtainable from:} Vilnius University Research Institute of Theoretical Physics and Astronomy,
A. Go\v{s}tauto 12, Vilnius, 2600, Lithuania.~~ E-mail: tomas@mserv.itpa.lt \\[10pt]
{\em Computer for which the library is designed and others on which it has been tested:} Pentium--based PC 155 MHz.\\
{\em Installations:} Vilnius University Research Institute of Theoretical Physics and Astronomy (Lithuania) \\
{\em Operating systems or monitors under which the new version has been tested:} LINUX 2.4.5 \\[10pt]
{\em Programming language used in the new version:} ANSI standard Fortran 90/95. \\[10pt]
{\em Memory required to execute with typical data:} 
     Memory requirements depend on the shell structure and the size of the wave function expansion  
     which is used to represent the atomic levels.  \\[10pt]
%
%
{\em No.\ of bits in a word:}  All real 
     variables are parametrized by a \texttt{selected kind parameter} and,  
     thus, can easily be adapted to any required precision as supported  
     by the compiler. Presently, the \texttt{kind} parameter is set to 
     double precision (two 32--bit words) in the module 
     \texttt{rabs\_{}constant}.\\[10pt] 
%
%
{\em No. of bytes in distributed program, including test data, etc.:} 
... \\[10pt]
{\em Distribution format:} compressed tar file \\[10pt]
{\em Keywords :} 
atomic structure, 
$LS$-coupling, $jj$-coupling,
$LS-jj$ transformation, 
optimal coupling scheme,
classification of atomic spectra.
\\[10pt]
{\em Nature of physical problem:}\\
The set of quantum numbers for the identification of atomic state functions (levels) and 
the classification of atomic spectra usually are chosen comparing the distributions of the
mixing coefficients in the bases corresponding to various coupling schemes.
The corresponding mixing coefficients usually are obtained 
diagonalizing the energy matrix 
in the basis of the wave functions of certain coupling.  
Once the secular equation is solved and the mixing coefficients in some coupling scheme are obtained
they can be used as a source to get corresponding coefficients for the bases of other coupling
schemes if the appropriate transformation matrix is known [1]. 
\\[10pt]
{\em Method of solution:} \\
For the transformation of the atomic state functions from $LS$- to $jj$-coupled basis we use the
recently obtained $LS-jj$ transformation matrices for the shells of equivalent electrons [2]
as well as the formula for the $LS-jj$ transformation matrices [3].
\\[10pt]
{\em Restrictions on the complexity of the problem:}\\
The program restricted to the configurations with $LS$-coupled shells with $l$=0, 1, 2 and 3.
\\[10pt]
{\em Unusual features of the program:} The \textsc{Lsjj} program is designed as a part of the coupling optimization package \textsc{Laris}.\\
\\[10pt]
{\em References:}
\begin{refnummer}

\item Z. Rudzikas, {\sl Theoretical Atomic Spectroscopy}
   (Cambridge Univ. Press, Cambridge, 1997).

\item G.~Gaigalas, T.~\v{Z}alandauskas and Z.~Rudzikas,
At.~Data  Nucl.~Data Tables, in print (2002).

\item G.~Gaigalas, Z.~Rudzikas and T.~\v{Z}alandauskas, 
Book of Abstracts of XI HCI Conference, 1-6 September, 2002, Caen, France, 48.







\end{refnummer}

\end{small}


\clearpage

\section*{LONG WRITE--UP}


\section{Introduction}

In investigating the energy spectra and other characteristics of atoms and ions, the
experimentalists most often use the $LS$-coupling scheme. There a shell of electrons is
made of the electron states having the same principal and angular quantum numbers, $n$ and
$l$. Such a shell is additionally characterized by the shell occupation number $N$ and the
resulting orbital and spin momenta, $L_i$ and $S_i$. 
When several shells are present, the
state in $LS$-coupling is additionally  characterized by the intermediate resulting momenta
$L_{ij}$ and $S_{ij}$, and by the total momenta $L$, $S$ and $J$. 
This coupling scheme is very popular in the theoretical atomic physics when atoms and ions 
are investigated in the non--relativistic approximation. 
This, in the first place, is related to the fact that
for the light atoms and ions such an approximation produces quite accurate results and the
agreement with the experiment is fairly good. 
Secondly, this coupling scheme is realistic for such systems. 
Therefore it is convenient for the identifying the energy levels. 
At the same time, this scheme is
convenient to investigate such processes as transitions, whereby the selection rules for
electronic transitions are both easily determined and reflect the real situation. 
But this coupling scheme is for a relatively small number of atoms and ions. 
The quantum numbers of this scheme are not accurate for highly charged ions and a number of other cases.
For the identification of atomic state functions (ASF) and classification of energy 
spectra the approximate 
quantum numbers are widely used. Usually such a classification relies on the
characteristics of the $LS-$ or $jj-$ coupling scheme.
But in a number of cases (for example ions of intermediate ionization degree) 
the choice of such classification is not trivial and implies the consideration 
of the expansions 
of atomic state functions in several bases of pure coupling wave functions.
Therefore it is convinient to have the possibility to transform the expansions of 
atomic state functions   
between various bases of pure coupling wave functions \cite{R:97}.

\medskip
  
The transformation matrices between the different bases of pure coupling wave functions   
were studied (\cite{Cowan:65}-\cite{GRZ:HCI02}), 
whereas the corresponding software were developed (\cite{Cowan:68,Ciplys:86,Zaland:2002,GF:2002}). 
Nevertheless until recently there were neither $LS-jj$ transformation matrices for the 
arbitrary number of open shells available, nor the corresponding software allowing one
to transform the ASFs resulted from the large 
scale calculations.
In addition, the software created earlier does not always allow to find the most suitable
coupling scheme, as sometimes it appears that no dominating scheme is present and several
schemes are equally suitable simultaneously. We present the investigation of the Fluorine 
isoelectronic sequence as an example~\cite{GKR:81}. 
In this case, several coupling schemes are
treated in the search for an optimal one. 
Those are $LS$, $LK$, $JK$ and $JJ$. In all those
schemes the $LS$ coupling is valid inside a shell. Using the analysis of the structure of
the weights of the ASF or the quantity R, one may notice that in this particular
isoelectronic sequence the ions do not have a dominating scheme, or the dominating
scheme does not stand out clearly among others. Therefore the above--mentioned methods have
been extended~\cite{GZR:01,GZR:02,GRZ:HCI02} in a way providing a possibility to change 
the $LS-$coupling scheme
into the $jj-$coupling one inside a shell, when necessary.
The recently obtained $LS-jj$ transformation matrices for the shell of equivalent electrons 
\cite{GZR:02} 
as well as the formulas for the $LS-jj$ transformation matrices 
\cite{GRZ:HCI02} 
enabled us to develop a computer code allowing to 
transform the ASFs from $LS$- to $jj$-coupling for 
all the configurations practically needed in atomic physics.
In the search for an optimal coupling scheme that would allow one to change the coupling
scheme inside a shell and to investigate the atoms and ions with ASFs containing open 
$f$-shells with arbitrary occupation numbers.

\medskip

To facilitate such a classification of the atomic levels and wave functions, we present here 
the program \textsc{Lsjj}. It supports the performance of the $LS-jj$ transformations. This program
may be of value as a separate module in solving the physical problems that also need this
transformation. 
However in this case one should bear in mind that the above--mentioned
transformation would provide the correct results only if the calculations are performed in
the framework of the Racah--Fano phase system~\cite{Gaigalas/RF:97}, and the coefficients 
of fractional parentage are defined via the quasispin formalism, as described 
in~\cite{Gaigalas/RF:98,Gaigalas/ZR:02}.

\medskip

A short explanation of the theoretical background is presented in Section 2. 
The explanation of the usage and the organization of the program with the description of main subroutines 
are presented in Section 3. 
The example of the usage of the program and the concluding remarks are presented in Sections 4 and 5 correspondingly.


\section{Theoretical background}

While investigating theoretically various characteristics of atoms and ions
(for example using multiconfiguration or configuration
interaction approaches) the atomic state functions are used which are either 
written in terms of a $LS-$coupled
\begin{eqnarray} 
\label{Psi-LS} 
   \ketm{\Psi_\tau (J^P)} & = & 
   \sum_{r} \, a_r^{\,(LS)} (\tau) \, \ketm{\gamma_r\ L_{r}S_{r}\ JP} 
\end{eqnarray} 
or $jj$-coupled basis 
\begin{eqnarray} 
\label{Psi-jj} 
   \ketm{\Psi_\tau (J^P)} & = & 
   \sum_{s} \, a_s^{\,(jj)} (\tau) \, \ketm{\gamma_s JP} \; ,
\end{eqnarray} 
where $\tau \,=\, 1,\ 2,\ ... $ enumerates the atomic states (of the given 
$J$ and parity $P$) and $\gamma_r,\ \gamma_s$ denote the sets of all additional quantum numbers  
of the $LS$- and $jj$-coupled configuration state functions (CSF) correspondingly. 

\medskip

When a pure $LS$--coupling exists in the atoms or ions under consideration, then 
$r$=1 in equation (\ref{Psi-LS}). When the $jj$--coupling exists, then 
$s$=1 in equation (\ref{Psi-jj}).  Therefore such an ASF is identified by an
array of quantum numbers $\gamma_1$.
However, these quantum numbers are exact only for the cases
of pure coupling schemes, which is more the exception than a rule.

\medskip

In the most general case the exact ASF is made up as an infinite sum in
(\ref{Psi-LS}) or (\ref{Psi-jj}). Therefore an ASF consists of an infinite
number of CSFs. In the real calculation one attempts to include terms
as many as possible in the sum having the maximum coefficients 
$a_r^{\,(LS)} (\tau)$ or $a_s^{\,(jj)} (\tau)$. But in this case it is not
clear what array $\gamma$ to ascribe in identifying the ASF. Bearing in mind
that the coefficient modulus squared (weight) 
$\left| a_{r}^{\,(LS)} (\tau) \right| ^{2}$ or 
$\left| a_{s}^{\,(jj)} (\tau) \right| ^{2}$ 
defines the probability for the atomic state to be in a state with the array of
quantum numbers $\gamma_{r} L_{r} S_{r}$ or $\gamma_{s}$, one can easily classify all the
energy levels by ascribing the characteristics of the maximum coefficient 
($a_{r_{max}}^{\,(LS)} (\tau)$ in $LS$ or 
$a_{s_{max}}^{\,(jj)} (\tau)$ in $jj$-coupling)
to the whole ASF.

\medskip

The situation becomes more complicated when 
the expansion (\ref{Psi-LS}) or (\ref{Psi-jj}) 
has no obviously dominant weight but several approximatelly equal ones.  
Most efficient way for such a classification is the transformation of the ASF into
another coupling scheme with the suitable distribution of weights 
(i.e. with only one dominant weight coefficient).  
For example, the classification of the spectra of complex atoms and ions 
using the intermediate quantum numbers of $LS$-coupling quite offen is problematical.

One of the possible solutions is to transform 
the function (\ref{Psi-LS}) into function (\ref{Psi-jj}).
Then the level is ascribed to the characteristics $\gamma_{r_{max}}$
that have the maximum weight $\left| a_{r_{max}}^{\,(jj)} (\tau) \right|$ 
at the transformation of CSFs $\ketm{\gamma_{r_{max}} \ JP}$, 
i.e. the intermediate quantum numbers of $jj$-coupling are used for the identification 
of the ASFs. 
Such transformations and identifications are the tasks of the \textsc{Lsjj} program 
presented in this paper.

\medskip

For practical purposes and especially for an efficient transformation
the program treats the CSF in a \textit{standard order}, 
i.e. it is assumed that the $LS$-coupled functions are couped consequently

\begin{eqnarray} 
\label{LS-csf}
   & & \hspace*{-0.8cm} 
\ketm{\gamma_r\ LS\ JP} \equiv
   \ketm{
    (...(((l_1^{\,N_1} \alpha_1 L_1 S_1, \,  
           l_2^{\,N_2} \alpha_2 L_2 S_2) L_{12}S_{12}, \, 
           l_3^{\,N_3} \alpha_3 L_3 S_3) L_{123}S_{123})...) LS\ J     
   }
\end{eqnarray} 

as well as the $jj$-coupled CSFs

\begin{eqnarray} 
\label{jj-csf} 
\lefteqn{
\ketm{\gamma_s JP}}
\nonumber \\[1ex]
   & \equiv &
   \ketm{(...(((((\tmm{\kappa}_1^{\tmm{N}_1} \tmm{\nu}_1 \tmm{J}_1,         \, 
             \tpm{\kappa}_1^{\tpm{N}_1} \tpm{\nu}_1 \tpm{J}_1) J_1,         \, 
             \tmm{\kappa}_2^{\tmm{N}_2} \tmm{\nu}_2 \tmm{J}_2) J^{\prime}_{12},\, 
             \tpm{\kappa}_2^{\tpm{N}_2} \tpm{\nu}_2 \tpm{J}_2) J_{12} 
             \tmm{\kappa}_3^{\tmm{N}_3} \tmm{\nu}_3 \tmm{J}_3) J^{\prime}_{123},\, 
             \tpm{\kappa}_3^{\tpm{N}_3} \tpm{\nu}_3 \tpm{J}_3) J_{123}...)J}, 
\end{eqnarray} 
where $\kappa$ is the relativistic (angular momentum) quantum number  
$\tmm{\kappa}_i  =  l_i$ and $\tpm{\kappa}_i  =  -l_i-1 $.  

\medskip

If both subshells with common $l_i$, i.e.\ $ \tmm{\kappa}_i$ and  
    $ \tpm{\kappa}_i$ appear in the expansion, these two subshells always 
    occur successively in the sequence 
    $ (\tmm{\kappa}_i^{\tmm{N}_i} \tmm{\nu}_i \tmm{J}_i, \, 
       \tpm{\kappa}_i^{\tpm{N}_i} \tpm{\nu}_i \tpm{J}_i) \ J_{i} $. 
    Formally, we can use this sequence even for the subshell states 
    with zero occupation if we interpret  
    $\ketm{\kappa^0 \nu=0\ J=0} \,\equiv\, 1$; in this case, the full 
    Clebsch--Gordan expansion remains valid due  
    to the orthonormality properties of the Clebsch--Gordan coefficients. 

\medskip

For the $LS-jj$ transformation of the configuration states
    we further assume in a \textit{standard order} that 
    $ l_1 \,=\, \tmm{\kappa}_1 \,=\, -(\tpm{\kappa_1}+1),\   
      l_2 \,=\, \tmm{\kappa}_2 \,=\, -(\tpm{\kappa_2}+1),\  ...$, i.e.\ 
    that the sequence of (sub--)shell states is the 
    \textit{same on both sides} of the transformation matrix. 

\medskip

The program presented in this paper is based on the methodics~\cite{{GZR:01}} where  
mixing coefficients $a_r^{\,(LS)} (\tau)$ 
are transformed to $a_s^{\,(jj)} (\tau)$ using the following relation:
%

\begin{eqnarray} 
\label{general-trans}
a_s^{\,(jj)} (\tau) = \sum_{r} \, 
\left\langle \gamma_s JP \vert \gamma_r\ L_{r}S_{r}\ JP \right\rangle
 \, a_r^{\,(LS)} (\tau) , 
\end{eqnarray} 

where $\left\langle \gamma_s JP \vert \gamma_r\ L_{r}S_{r}\ JP \right\rangle$ is the $LS-jj$ transformation matrix.

\medskip

In general, the performing of the transformations (\ref{general-trans}) with the matrix  
$\left\langle \gamma_s JP \vert \gamma_r\ L_{r}S_{r}\ JP \right\rangle$ requires 
the $LS-jj$ transformation matrices for the shell of equivalent electrons \cite{GRZ:HCI02}.
The numerical values of the matrices for the shell of equivalent electrons are presented 
in \cite{GZR:02}, nevertheless 
it is necessary to ensure the compatibility of the coefficients of fractional 
parentage (CFP) used to form (\ref{LS-csf}) and (\ref{jj-csf}) with the ones used in the 
transformation matrices.
In this program the usage of the CFP obtained from the reduced coefficients (subcoefficients) 
of fractional parentage (RCFP) is assumed \cite{R:97} 
(such CFPs were used for the calculation of $LS-jj$ matrix elements). 
The program is based on quasisipin formalism and therefore the appropriate data are required. 
The data could be generated using the popular package \cite{MCHF1978, Fbook} with 
the angular parts \cite{G:02}.

\medskip



\section{Description of the Program}

The program is written in Fortran 90 programming language. 
The main part of the program consists of several subroutines collected into the 
module \texttt{rabs\_lsjj} (file \texttt{rabs\_lsjj.f90}). 
The program is designed in the way similar to \textsc{Ratip} 
\cite{Ratip} package. 
Some of the \textsc{Ratip} features are used and the overall program includes the additional modules 
\texttt{rabs\_angular},
\texttt{rabs\_constant},
\texttt{rabs\_csl},
\texttt{rabs\_determinant},
\texttt{rabs\_dirac\_orbital},
\texttt{rabs\_naglib},
\texttt{rabs\_nucleus},
\texttt{rabs\_rcfp},
\texttt{rabs\_utilities},
\texttt{rabs\_lsj},
\texttt{rabs\_lsj\_data}
\cite{GZF:2003, FritzscheFG:02} (the corresponding files named as the \texttt{module\_name.f90}).

\subsection{Installation of the program}
\label{install}

The program was designed and tested on the PC with the Linux OS. 
Nevertheless it can be easily adopted to any Unix-Linux based platforms 
(including \textsc{IBM RS}/6000, \textsc{Sun OS}).    
The installation of the program is similar to \textsc{Ratip} components. 
First, the script \texttt{make-environment} which declares the global variables 
for the compilation and linkage of the program should be run via statement 
\texttt{source make-environment}. 
Then, the make-file (\texttt{make-lsjj} in this case) should be executed via comand 
\texttt{make -f make-lsjj} (both files can be obtained from the authors together 
with other program files). 
Then the executable \texttt{xlsjj} is generated.

\subsection{Interactive control and output of the program}
\label{interactive}

The principled scheme of the execution of the program is presented in Figure~\ref{fig:execution}.
After the initialization of several data arrays at the beginning of the execution 
(subroutine \texttt{lsj\_initialization\_LS\_jj()} \cite{GZF:2003}), the control is  
taken by the procedure \texttt{lsjj\_control\_transformation()}, which reads 
and interprets all input data, carries out the transformation for all selected
levels, and finally prints the results to screen. 
The example of an interactive dialogue performed through the execution is presented in 
Figure~\ref{fig:interactive}.

\subsubsection{Input}
\label{input}

The input of data in this version of the program is performed by the subroutine
\texttt{lsjj\_load\_csfs\_from\_MCHF\_files()} called from the 
subroutine \texttt{lsjj\_control\_transformation()}.
The input data is assumed to be in the same format as the \textsc{Mchf} \cite{MCHF1978, Fbook} output data and can 
be generated with this package with the angular parts \cite{G:02} 
(to ensure the usage of the CFP defined via the quasispin formalism). 
Two files, one with the CSF basis (configuration list file \texttt{.inp}) 
and the other with the ASFs expansions (mixing coefficients file \texttt{.j}) should be provided.
The configuration list file may be generated using the \textsc{Genclf} and the mixing coefficients file - 
using the \textsc{Ci} program from the \textsc{Mchf} package \cite{MCHF1978}. 
The fragments of the input files are presented in Figures \ref{fig:inp}, \ref{fig:j}.
The detailed specification of the format of these files is avialable at the description of the popular 
\textsc{Mchf} package (mentioned above) therefore we will not go into details.

\subsubsection{Output}
\label{output}

The output of the program, namely the $jj$-coupled CSFs and the expansions of the selected 
ASFs are performed either directly from the subroutine \texttt{lsjj\_control\_transformation()}
or from the specific data output subroutines called from it. 

The direct "onscreen" output includes the quantum numbers and weights of 
several CSFs with the greatest weights as well as 
the sum of the squares of the mixing coefficients 
(in the output of the program denoted as \texttt{Total sum of weights}), 
serving as the numerical criteria of the accuracy of the expansions \ref{Psi-LS}, \ref{Psi-jj}.

The procedures 
\texttt{lsjj\_print\_configuration\_scheme\_jj()}, \texttt{lsjj\_print\_coefficients\_jj()}
also print a \textit{full expansions} of all ASFs in   
$jj$-coupled basis to the files \texttt{lsjj-out-csfs.jj} and \texttt{lsjj-out-coeff.jj} 
in the manner similar to the $LS$-coupled data.
The fragments of the output files are presented in Figures \ref{fig:jj-csfs}, \ref{fig:jj-coef}.


\subsection{Description of the module \texttt{rabs\_lsjj}}
\label{modlsjj}

In this section we will introduce the main procedures of the module and explain some details of their 
usage because the main module of the program \texttt{rabs\_lsjj} is assumed to be used within 
the \textsc{Lsjj} as well as in other programs 
(for example in other programs of the coupling optimization package \textsc{Laris}).   
The module collects the procedures and functions in an alphabetic order; they are briefly 
explained in the header of the module. 
Further information about the methods, data structures, etc.\ and their implementation 
can be found in many in--line comments in the headers of the individual procedures/functions or 
directly inside the source code.
The names of the subroutines/functions in the module begin with the "\texttt{lsjj\_}" and are followed 
by the rest of the names explaining the purpose of the procedure.
Table 1 lists the procedures and functions grouped according to their purpose. 
The detailed description of the subroutines/functions is presented below.

\begin{table} 
\begin{small} 
 
{\bf Table 1} 
\hspace{0.2cm} 
{\rm  
The subroutines and functions of the \texttt{rabs\_lsjj} module.
} 
\begin{center} 
\begin{tabular}{l c p{10.6cm}} \\[-0.3cm]   \hline \hline           \\[-0.3cm] 
   General purpose   & &  
                                                                    \\[0.1cm] 
      & &  
      \texttt{lsjj\_control\_transformation()}                        \\[0.1cm] 
      & &  
      \texttt{lsjj\_load\_csfs\_from\_MCHF\_files()}                  \\[0.1cm] 
      & &  
      \texttt{lsjj\_form\_csf\_basis\_jj()}                           \\[0.1cm] 
      & &  
      \texttt{lsjj\_transformation\_ASF()}                            \\[0.1cm] 
      & &  
      \texttt{lsjj\_get\_subshell\_term\_jj()}                        \\[0.1cm] 
   Output facilities       & &  
                                                                    \\[0.1cm] 
      & &  
         \texttt{lsjj\_print\_MCHF\_data()}                           \\[0.1cm] 
      & &  
         \texttt{lsjj\_print\_csfs\_LS\_MCHF()}                       \\[0.1cm] 
      & &  
         \texttt{lsjj\_print\_coefficients\_LS\_MCHF()}               \\[0.1cm] 
      & &  
         \texttt{lsjj\_print\_single\_configuration\_jj2()}           \\[0.1cm] 
      & &  
         \texttt{lsjj\_print\_coefficients\_jj()}                     \\[0.1cm] 
      & &  
         \texttt{lsjj\_print\_configuration\_scheme\_jj()}            \\[0.1cm] 
      & &  
         \texttt{lsjj\_print\_detailed\_info\_csf\_set\_LS()}         \\[0.1cm] 
    Utilities   & &  
                                                                    \\[0.1cm] 
      & &  
          \texttt{lsjj\_lval()}                                       \\[0.1cm] 
      & &  
           \texttt{lsjj\_lvalr()}                                     \\[0.1cm] 
      & &  
          \texttt{lsjj\_largest\_coeff()}                             \\[0.1cm] 
      & &  
          \texttt{lsjj\_deallocate\_asf\_basis\_jj()}                 \\[0.1cm] 
   \hline \hline  
\end{tabular} 
\end{center} 
\end{small} 
 
\vspace{0.7cm} 
\end{table}


\subsubsection*{The subroutine \texttt{lsjj\_control\_transformation()}}

The main subroutine is controling the flow of the execution. 
The subroutine performs an interactive dialogue with the user, 
reads the specification parameters for the calculation, opens and closes
input and output streams, calls other subroutines of data input, the proceeding and the output. 
No direct input arguments should be specified.
All the dinamically allocated arrays are deallocated at the end of the execution of the subroutine.

\subsubsection*{The subroutine \texttt{lsjj\_load\_csfs\_from\_MCHF\_files()}}

This subroutine loads the data on ASFs and corresponding expansions in terms of $LS$-coupled CSFs from the files. 
The input files are assumed to be formated like MCHF output files \cite{Fbook}. 
Two input arguments \texttt{stream\_nr and stream\_nr\_j} specify the input streams for the configuration 
list file and for mixing coefficient file. 
The output argument \texttt{ierr} is designed for error handling and is assigned to 0 in case of failure.
The data read are placed to the public variable \texttt{asf\_set\_LS\%csf\_set\_LS}.

\subsubsection*{The subroutine \texttt{lsjj\_form\_csf\_basis\_jj()}}

This subroutine generates the $jj$-coupled CSFs corresponding to the $LS$-coupled ones stored 
in the public variable \texttt{asf\_set\_LS\%csf\_set\_LS}. 
No direct input arguments should be specified.
The data generated are placed to the public variable \texttt{asf\_set\%csf\_set}.
The algorithm of the generation is designed in a way that the subroutine can deal with the 
wave functions of the configurations with the arbitrary number of open shells. 
The algorithm is implemented using the utility of recursive subroutines. 
Two internal recursive subroutines are used. 
The first one \texttt{lsjj\_form\_csf\_basis\_jj\_fill\_occupations()} is meant for 
the definition of the number of $jj$-coupled shells (and occupation numbers $\tmm{N}_i, \tpm{N}_i$) 
for the given $LS$ coupled shell $l_{i}^{N_i}$.
And the second one \texttt{lsjj\_form\_csf\_basis\_jj\_job\_count()} is for the the calculation of 
the number of CSFs in $jj$-coupling and corresponding intermediate quantum numbers 
$J^{\prime}_{i..j}, J_{i..j}$ (\ref{jj-csf}).

\subsubsection*{The subroutine \texttt{lsjj\_transformation\_ASF()}}

The subroutine transforms the atomic state function, represented 
in a $LS$-coupling CSF basis into a basis of $jj$-coupling CSF.
Namely the expansion coefficients $a_s^{\,(jj)} (\tau)$ are calculated and placed to the 
array \texttt{asf\_set\%asf(level\_LS)\%eigenvector}.
The input argument \texttt{level\_LS} specifies a serial number of the state to be considered.
The $LS-jj$ matrix elements $\left\langle \gamma_s JP \vert \gamma_r\ LS\ JP \right\rangle$ 
for the transformation (\ref{general-trans}) are taken from module 
\texttt{rabs\_lsj\_data} via the subroutine \texttt{lsj\_transformation\_LS\_jj\_general()} 
\cite{GZF:2003}.

\subsubsection*{The subroutine \texttt{lsjj\_get\_subshell\_term\_jj()}}

This procedure returns all allowed subshell terms (j, w, Q, J) for given $j^N$ 
which must be 1/2, 3/2, 5/2, 7/2 or 9/2. 
Two input arguments namely \texttt{j\_shell} and \texttt{N} define the value 
of the angular momentum and the occupation. 
The number of corresponding terms as well as their quantum numbers are given by 
the output arguments \texttt{number} and \texttt{jj} correspondingly.


\bigskip

There are a number of subroutines for the data output in various forms. 
All the names for all subroutines begin with the \texttt{lsjj\_print\_}
and are followed by the rest of the names explaining the content and type of the output.

\medskip

{\bf The subroutine \texttt{lsjj\_print\_MCHF\_data()} }

\smallskip

The subroutine prints the data in the manner of MCHF program.
The input arguments
\texttt{streamnr\_inp} and \texttt{streamnr\_j} specify the output streams for the 
MCHF mixing coefficients (\texttt{.j}) and configuration list (\texttt{.inp}) files correspondingly.
The argument \texttt{asf\_set\_LS} specifies the variable where the $LS$-coupled data are stored.
The subroutine first generates the temporary set of CSFs  
not coupled to $J$ (as it is in MCHF files), 
prescribes the mixing coefficients to this set of CSFs and stores this data in temporary
variable \texttt{asf\_set\_LS\_MCHF}. 
Then the subroutine calls \texttt{lsjj\_print\_csfs\_LS\_MCHF()} and 
\texttt{lsjj\_print\_coefficients\_LS\_MCHF()} for the output of CSFs and ASFs with 
the corresponding mixing coefficients from this temporary variable.

\medskip

{\bf The subroutine \texttt{lsjj\_print\_csfs\_LS\_MCHF()} }

\smallskip

The subroutine prints the configuration state functions to the file in the form similar to MCHF configuration 
list file (\texttt{.inp}).
The argument \texttt{streamnr} specifies the output stream and the \texttt{csf\_set\_LS}
specifies the CSFs in $LS$-coupling (not coupled to $J$). 

\medskip

{\bf The subroutine \texttt{lsjj\_print\_coefficients\_LS\_MCHF()} }

\smallskip

The subroutine prints the data on ASFs with expansion coefficients to the file in the form similar to 
MCHF mixing coefficients \texttt{.j} file.
The argument \texttt{streamnr} specifies the output stream and the \texttt{asf\_set\_LS}
specifies the ASFs and the mixing coefficients in $LS$-coupling 
(for the list of CSFs not coupled to $J$).

\medskip

{\bf The subroutine \texttt{lsjj\_print\_single\_configuration\_jj2()} }

\smallskip

The subroutine prints all information about a single CSF in jj-coupling.
The input argument \texttt{stream} specifies the output stream,
the \texttt{csf\_set} specifies the variable of type \texttt{csf\_basis} where 
information on $jj$-coupled CSFs is stored  and \texttt{csf\_number} the serial 
number of CSF to be printed. 
The subroutine is similar to the \texttt{lsjj\_print\_single\_configuration\_jj()} 
defined in module \texttt{rabs\_lsj}.

\medskip

{\bf The subroutine \texttt{lsjj\_print\_coefficients\_jj()} }

\smallskip

The subroutine prints expansion coefficients in $jj$-coupling to the file in the form similar to MCHF mixing 
coefficients (\texttt{.j}) file.
The input argument \texttt{streamnr} specifies the output stream and the \texttt{asf\_set} 
specifies the variable where the $jj$-coupled data are stored.

\medskip

{\bf The subroutine \texttt{lsjj\_print\_configuration\_scheme\_jj()} }

\smallskip

The subroutine prints information about the CSFs in $jj$-coupling.
The input argument \texttt{stream} specifies the output stream and the \texttt{csf\_set} 
specifies variable where the data on $jj$-coupled CSFs are stored.

\medskip

{\bf The subroutine \texttt{lsjj\_print\_detailed\_info\_csf\_set\_LS()} }

\smallskip

The subroutine prints detailed info about the $LS$-coupled CSFs. 
The input argument \texttt{stream} specifies the output stream and the \texttt{csf\_set\_LS} 
specifies variable where the data on $LS$-coupled CSFs are stored.

\medskip

The module \texttt{rabs\_lsjj} contains a number of auxiliary subroutines and functions.

\medskip

{\bf The function \texttt{lsjj\_lval()} }

\smallskip

The function returns the integer value of an orbital quantum number given as a character.
The range of an argument \texttt{l} should be from s to q (or from S to Q), 
i.e. the corresponding angular momentum $l$ from 0 to 12.

\medskip

{\bf The function \texttt{lsjj\_lvalr()} }

\smallskip

The function returns the character value of an orbital quantum number given as an integer.
The range of an argument \texttt{l} should be from 1 to 12.
The subroutine is inverse to the \texttt{lsjj\_lval()}.

\medskip

{\bf The function \texttt{lsjj\_largest\_coeff()} }

\smallskip

The function defines the serial number of the largest element in an array.
The argument \texttt{array} specifies the dinamically allocated array
and \texttt{size} - it's size.

\medskip

{\bf The subroutine \texttt{lsjj\_deallocate\_asf\_basis\_jj()} }

\smallskip

The subroutine deallocates the dinamically allocated data arrays within the variable \texttt{asf\_set}
(i.e. the $jj$-coupled data). 
The variable is defined in the module \texttt{rabs\_lsj}.
No direct input arguments should be specified.




\section{Example}

Here we present the example of the transformation of the atomic state function 
and classification in terms of quantum numbers of $LS$- and $jj$-coupling schemes.
The example considers the excited state of Ne II. 
The interactive dialogue performed in the process of the execution of the program 
is presented in Figure~\ref{fig:example}. 

\medskip

First the names of configuration list file (\texttt{Ne.inp}) and mixing coefficients file \texttt{Ne.j}) 
are specified 
(the fragments of the input files presented in Figures \ref{fig:inp} and \ref{fig:j}).
Then the program proceeds the input and asks to specify the ASFs to consider.

\medskip

After the specification of the ASFs the corresponding $jj$-coupled CSF basis is formed (144 CSFs in this case) 
and the mixing coefficients $a_{s}^{\,(jj)} (\tau)$ of the MCHF expansion (\ref{Psi-jj}) are calculated.
The resultant CSF basis and mixing coefficients of the specified ASF are printed to the files 
\texttt{lsjj-out-csfs.jj} and \texttt{lsjj-out-coeff.jj} correspondingly. 
The fragments of the output files are presented in Figures \ref{fig:jj-csfs} and \ref{fig:jj-coef}.
In addition the $LS$- and $jj$-coupled CSFs with the greatest weights (squares of the mixing coefficients) 
useful for the identification and classification of ASFs under the consideration presented interactively.
In our case the considered ASF may be identified by means of the quantum numbers of the $LS$-coupled CSF 
$\ketm{(((2s^2){}^{1}S, (2{p}^4){}^{3}P)~{}^{3}P,(3{s}^1){}^{2}S) ~ {}^{4}P_{\frac{5}{2}}}$ 
or by means of the quantum numbers of $jj$-coupled CSF 
$\ketm{((((2s^2)0, (2{p-}^2)0)~0, (2{p}^2)2)~2, (3s^1)\frac{1}{2}) ~ \frac{5}{2}}$.

\medskip

The sums of the squares of the mixing coefficients in $LS$- and $jj$-couplings 
equal to 1 with the precision of the input coefficients in $LS$-coupling (0.0000001). 
The fact that the sums in $LS$- and $jj$-couplings match up to the thirteenth digit reveals 
the accuracy of the performation of the transformation (\ref{general-trans}).

\bigskip

\section{Conclusions and outlook}

The program \textsc{Lsjj} enables us to transform the multiconfigurational expansions 
of the atomic state functions from $LS$- to $jj$-coupled bases. 
The input data in the current version of the program are supposed to be in the form of 
MCHF \cite{Fbook} output files, but the structure of the program enables to add more 
data input interfaces easily.  
The module \texttt{rabs\_lsjj} together with the earlier constructed \texttt{rabs\_lsj} 
\cite{GZF:2003} extends the possibility of comparison of ASFs expansions in various 
bases.
This extent combined together with the program for the changing coupling scheme between 
the $LS$-coupled shells of equivalent electrons and evaluation of the suitability of the 
intermediate quantum numbers for the identification and classification of atomic spectra 
under consideration \cite{Zaland:2002} will serve as a tool for the search of the optimal 
set of quantum numbers for the classification of spectra of atoms and ions.

\medskip

\bigskip

Program is obtainable from Vilnius University Research Institute of Theoretical Physic and Astronomy,
A. Go\v stauto 12, Vilnius, 2600, LITHUANIA.~~ E-mail: tomas@mserv.itpa.lt.

\newpage

PROGRAMA ATOM\c{U} B\={U}SEN\c{U} FUNKCIJOMS TRANSFORMUOTI I\v{S} $LS$ \c{I} $jj$ RY\v{S}IO 
BANGINI\c{U} FUNKCIJ\c{U} BAZ\c{E}

\medskip

T. \v{Z}alandauskas ir G. Gaigalas

\medskip
\medskip
\medskip

Santrauka

\medskip
Eksperimenti\v{s}kai gaunami atom\c{u} ir jon\c{u} energetiniai spektrai paprastai identifikuojami ir 
klasifikuojami naudojant LS ry\v{s}io schemos kvantinius skai\v{c}ius.
\v{S}i ry\v{s}io schema yra labai populiari ir teorin\.{e}je atomo fizikoje, 
kai atomai ir jonai yra nagrin\.{e}jami nereliatyvistiniame artutinume. 
Bet $LS$ ry\v{s}io kvantiniai skai\v{c}iai 
da\v{z}nai n\.{e}ra tinkami sunki\c{u} atom\c{u} bei didelio ionizacijos laipsnio jon\c{u} b\={u}senoms  
apibudinti. 
Vienareik\v{s}miam bei tiksliam toki\c{u} atom\c{u} bei jon\c{u} b\={u}sen\c{u} identifikavimui bei 
spektr\c{u} klasifikavikavimui geriau tinka kit\c{u} ry\v{s}io shem\c{u} (da\v{z}nai $jj$ ry\v{s}io) 
kvantiniai skai\v{c}iai. 

Neseniai gautos $LS$-$jj$ transformacijos matricos ekvivalentini\c{u} elektron\c{u} sluoksniams 
(\c{i}skaitant $f$ elektron\c{u} sluoksnius su bet kokiu u\v{z}pildymo skai\v{c}iumi) 
bei $LS$-$jj$ transformacijos matric\c{u} i\v{s}rai\v{s}kos konfig\={u}racijoms su bet kokiu atvir\c{u} sluoksni\c{u} 
skai\v{c}iumi \c{i}galino mus sukurti program\c{a}, skirt\c{a} atom\c{u} bei jon\c{u} b\={u}sen\c{u} 
funkcij\c{u} transformavimui i\v{s} $LS$ \c{i} $jj$ ry\v{s}io bangini\c{u} funkcij\c{u} baz\c{e}. 
Tokio pob\={u}d\v{z}io program\c{u}, leid\v{z}ian\v{c}i\c{u} keisti ry\v{s}io tip\c{a} tiek ekvivalentini\c{u} 
elektron\c{u} sluoksni\c{u} viduje (\c{i}skaitant $f$ elektron\c{u} sluoksnius), 
tiek ir tarp sluoksni\c{u}, iki \v{s}iol nebuvo. 

Programa \c{i}galina atlikti atom\c{u} ir jon\c{u} spektr\c{u} identifikavim\c{a} ir klasifikacij\c{a} 
$LS$ ir $jj$ ry\v{s}io kvantiniais skai\v{c}iais,
praple\v{c}ia kvantini\c{u} skai\v{c}i\c{u} rinkinio, optimalaus atomini\c{u} spektr\c{u} 
identifikavimui ir klasifikavimui, paie\v{s}kos galimybes.

\newpage
\pagebreak


%
{\renewcommand\baselinestretch{0.85} 
\begin{figure}
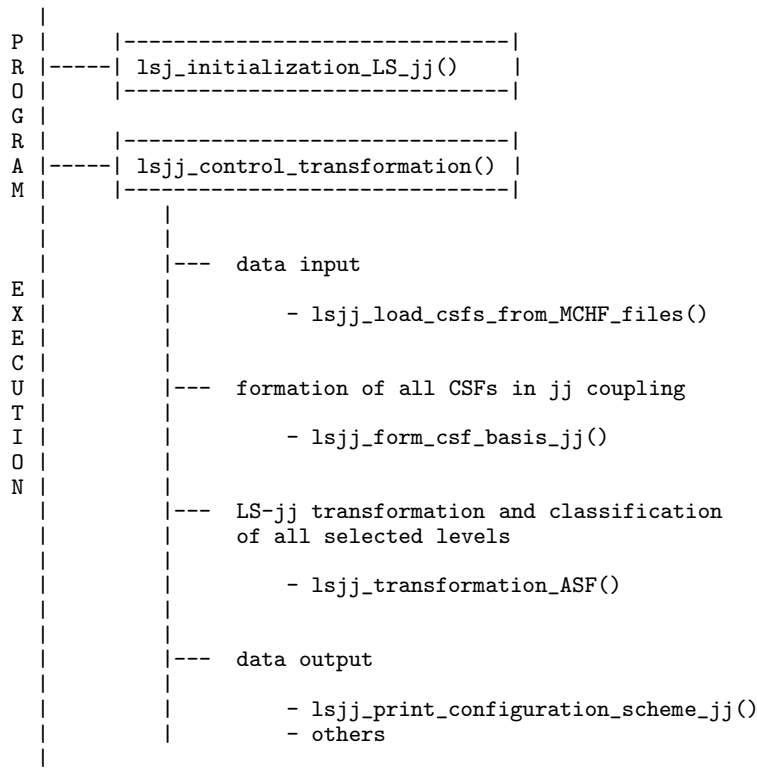
 
\begin{small} 
\begin{verbatim} 
                 | 
               P |     |-------------------------------| 
               R |-----| lsj_initialization_LS_jj()    | 
               O |     |-------------------------------| 
               G | 
               R |     |-------------------------------| 
               A |-----| lsjj_control_transformation() | 
               M |     |-------------------------------| 
                 |         | 
                 |         | 
                 |         |---  data input 
               E |         | 
               X |         |         - lsjj_load_csfs_from_MCHF_files() 
               E |         |
               C |         |
               U |         |---  formation of all CSFs in jj coupling
               T |         | 
               I |         |         - lsjj_form_csf_basis_jj() 
               O |         |
               N |         | 
                 |         |---  LS-jj transformation and classification 
                 |         |     of all selected levels
                 |         | 
                 |         |         - lsjj_transformation_ASF()
                 |         | 
                 |         | 
                 |         |---  data output 
                 |         | 
                 |         |         - lsjj_print_configuration_scheme_jj()
                 |         |         - others
                 |         
\end{verbatim} 
\end{small} 
\caption{\small Schematic flow chart of the \textsc{Lsjj} program.} 
\label{fig:execution} 
\end{figure} 
}

\begin{figure}
\begin{small}
\begin{verbatim}
  Program for transformation of atomic states from LS to jj coupling
  (C) Copyright by T Zalandauskas and G Gaigalas Vilnius (2003).
 Transform one or several ASF from a MCHF calculation into a     
 jj-coupled CSF basis. The transformation starts from the given  
 cfg.inp and .j files and is carried out for the n leading CSF in
 the LS-coupled basis; the new representation in the jj basis is 
 printed as in a standard GRASP92 computation.                   
  
 Enter the name of the MCHF configuration list file:
>
 Enter the name of corresponding .j mixing coefficient file:
>
 load_cfg_from_MCHF_file ... 
  number of core shells  =  ...
  total number of shells =  ...
  number of csfs LS      =  ...
  reordered 
  there are  ... nonrelativistic CSFs (before couplin to J);
  now couple them to J ...
  now load mixing coefficients and form asf_set_LS... 
  end reading j
  there are  ...  atomic state functions
  there are  ...  nonrelativistic CSFs (coupled to J)
  ... load complete.
 Maximum number of considered ASF is: ...
 Enter the level numbers of the ASF which are to be transformed,
  e.g. 1 3 4  7 - 20  48  69 - 85 :
>
\end{verbatim}
\end{small}
\caption{\small Interactive dialogue of \textsc{Lsjj} program} 
\label{fig:interactive} 
\end{figure}


\begin{figure}
\begin{small}
\begin{verbatim}
  1s 
  2s( 2)  2p( 4)  3s( 1)
     1S0     1S0     2S1     1S      2S  
  2s( 2)  2p( 4)  3s( 1)
     1S0     3P2     2S1     3P      2P  
  2s( 2)  2p( 4)  3s( 1)
     1S0     3P2     2S1     3P      4P  
  2s( 2)  2p( 4)  3s( 1)
     1S0     1D2     2S1     1D      2D  
  2s( 2)  2p( 4)  3d( 1)
     1S0     1S0     2D1     1S      2D  
  2s( 2)  2p( 4)  3d( 1)
     1S0     3P2     2D1     3P      2P  
  2s( 2)  2p( 4)  3d( 1)
     1S0     3P2     2D1     3P      4P  
  2s( 2)  2p( 4)  3d( 1)
     1S0     3P2     2D1     3P      2D  
  2s( 2)  2p( 4)  3d( 1)
     1S0     3P2     2D1     3P      4D  
  2s( 2)  2p( 4)  3d( 1)
     1S0     3P2     2D1     3P      4F  
  2s( 2)  2p( 4)  3d( 1)
     1S0     1D2     2D1     1D      2P  
  2s( 2)  2p( 4)  3d( 1)
     1S0     1D2     2D1     1D      2D  
  2s( 2)  2p( 4)  3d( 1)
...
\end{verbatim}
\end{small}
\caption{\small Fragment of MCHF configuration list file \texttt{Ne.inp}} 
\label{fig:inp} 
\end{figure}

\begin{figure}
\begin{small}
\begin{verbatim}
  Ne      Z =  10.0  N =   9   NCFG = 224



  2*J =    5  NUMBER = 1

     3   -126.95806136
 0.0000000 0.0000000 0.9951298 0.0130359 0.0003521 0.0000000 0.0237319
 0.0096379 -.0179847 -.0159071 0.0000000 -.0002664 0.0000000 0.0177844
 0.0000002 0.0000000 0.0000000 0.0000000 0.0000000 0.0000000 0.0000000
 0.0575015 -.0000647 -.0005302 0.0000000 0.0001279 0.0000000 0.0000000
...
\end{verbatim}
\end{small}
\caption{\small Fragment of MCHF mixing coefficients file \texttt{Ne.j}} 
\label{fig:j} 
\end{figure}


\begin{figure}
\begin{small}
\begin{verbatim}
  
 The current configuration scheme with 144 CSF in jj coupling
                              is defined as follows:
  
 Number of electrons:  9
  
 Number of (relativistic) subshells: 9
 Core shells:
  1s 
 Peel shells:
  2s     2p-    2p     3s     3p-    3p     3d-    3d 
 144  CSF(s):
            1) 2s ( 2)   2p-( 1)   2p ( 3)   3s ( 1)
                            1/2       3/2       1/2 
                               1/2        2         5/2  
            2) 2s ( 2)   2p-( 2)   2p ( 2)   3s ( 1)
                                      2         1/2 
                                          2         5/2  
            3) 2s ( 2)   2p ( 4)   3d ( 1)
                                      5/2 
                                         5/2  
            4) 2s ( 2)   2p-( 1)   2p ( 3)   3d ( 1)
                            1/2       3/2       5/2 
                               1/2        1         5/2  
            5) 2s ( 2)   2p-( 1)   2p ( 3)   3d ( 1)
                            1/2       3/2       5/2 
                               1/2        2         5/2  
            6) 2s ( 2)   2p-( 1)   2p ( 3)   3d-( 1)
                            1/2       3/2       3/2 
                               1/2        1         5/2  
...
\end{verbatim}
\end{small}
\caption{\small Fragment of output file \texttt{lsjj-out-csfs.jj} of $jj$ coupled CSFs} 
\label{fig:jj-csfs} 
\end{figure}

\begin{figure}
\begin{small}
\begin{verbatim}
     1   -126.95806136
  0.5638947  0.8200464  -.0140585  0.0254398  0.0123155  -.0067862  -.0061464
  0.0103721  0.0170042  -.0084857  0.0230858  -.0169735  0.0440080  -.0234300
  0.0355732  0.0514159  -.0002857  -.0018696  0.0043991  0.0004125  -.0016268
  0.0049082  0.0044622  -.0028572  -.0023440  0.0025700  0.0059582  0.0018875
...
\end{verbatim}
\end{small}
\caption{\small Fragment of output file \texttt{lsjj-out-coeff.jj} of the ASFs expansions in $jj$-coupling} 
\label{fig:jj-coef} 
\end{figure}


\begin{figure}
\begin{small}
\begin{verbatim}
.
.
 Enter the name of the MCHF configuration list file:
Ne.inp
 Enter the name of corresponding .j mixing coefficient file:
Ne.j
 load_cfg_from_MCHF_file ... 
  number of core shells  =  1
  total number of shells =  6
  number of csfs LS      =  224
  there are  224 nonrelativistic CSFs (before coupling to J);
  now couple them to J ...
  now load mixing coefficients and form asf_set_LS... 
  end reading j
  there are  1  atomic state functions
  there are  144  nonrelativistic CSFs (coupled to J)  
  ... load complete.
 Maximum number of considered ASF is: 1
 Enter the level numbers of the ASF which are to be transformed,
  e.g. 1 3 4  7 - 20  48  69 - 85 :
1
 number_of_levels, levels(:) =  1 1
  start formation of CSFs in jj coupling...
    number of shells in jj coupling =  9
    number of core shells in jj coupling =  1
    number of configuration state functions in jj coupling =  144
  finish formation of CSFs in jj coupling...
  
 Weights of major contributors to ASF in LS-coupling:
  Level  J Parity      CSF contributions
  
    1  5/2   +      0.99028 of    1    0.00391 of   15    0.00331 of   11    
  
 Definition of leading CSF:
         1)     2s( 2)   2p( 4)   3s( 1)
                   1S0      3P2      2S1      3P       4P       5/2 
  
  Total sum of weights is:  0.9999999614146498
  
  
 Weights of major contributors to ASF in jj-coupling:
  Level  J Parity      CSF contributions
    1  5/2   +      0.67248 of    2    0.31798 of    1    0.00264 of   16

  Total sum of weights is:  0.9999999614146532
  
 Definition of leading CSF:
  
            1) 2s ( 2)   2p-( 1)   2p ( 3)   3s ( 1)
                            1/2       3/2       1/2 
                               1/2        2         5/2  
            2) 2s ( 2)   2p-( 2)   2p ( 2)   3s ( 1)
                                      2         1/2 
                                          2         5/2  
           16) 2s ( 1)   2p-( 2)   2p ( 2)   3s ( 2)
                  1/2                 2             
                               1/2        5/2       5/2  
  
 LS-jj complete ... .
\end{verbatim}
\end{small}
\caption{\small Interactive dialogue of \textsc{Lsjj} program for the transformation of state function of excited Ne} 
\label{fig:example} 
\end{figure}

\end{document}